\documentclass[aps,pre,groupedaddress,twocolumn,10pt]{revtex4-1}
\usepackage{graphicx}
\usepackage{epstopdf}
\usepackage[english]{babel}
\usepackage[T1]{fontenc}
\usepackage{verbatim}
\usepackage{float}
\usepackage{makecell}
\usepackage{color}
\usepackage{multirow}
\usepackage{slashbox}
\usepackage{booktabs}
\usepackage{amsmath}
\usepackage{amssymb}
\usepackage{amsthm}
\usepackage{bbm}
\usepackage{hyperref}
\usepackage[T1]{fontenc}
\usepackage[usenames,dvipsnames]{xcolor}
\hypersetup{colorlinks=true, linkcolor=blue, urlcolor=blue!50!black, citecolor=blue!50!black}

\usepackage[normalem]{ulem}

\renewcommand{\imath}[0]{\mathsf{i}}

\usepackage[nodayofweek,level]{datetime}

\begin{document}

\title{How does an external electric field trigger the Cassie-Baxter-Wenzel wetting transition on a textured surface?}

\author{Ke Xiao$^{\dag}$, Xi Chen$^{\dag}$, and Chen-Xu Wu}
\email{cxwu@xmu.edu.cn}
\affiliation{Fujian Provincial Key Laboratory for Soft Functional Materials Research, Department of Physics, College of Physical Science and Technology, Xiamen University, Xiamen 361005, People's Republic of China}

\begin{abstract}
Understanding the critical condition and mechanism of the droplet wetting transition between Cassie-Baxter state and Wenzel state triggered by an external electric field is of considerable importance because of its numerous applications in industry and engineering.
However, such a wetting transition on a patterned surface is still not fully understood, e.g., the effects of electro-wetting number, geometry of the patterned surfaces, and droplet volume on the transition have not been systematically investigated.
In this paper, we propose a theoretical model for the Cassie-Baxter-Wenzel wetting transition triggered by applying an external voltage on a droplet placed on a mirco-pillared surface or a porous substrate. It is found that the transition is realized by lowering the energy barrier created by the intermediate composite state considerably, which enables the droplet to cross the energy barrier and complete the transition process. Our calculations also indicate that for fixed droplet volume, the critical electrowetting number (voltage) will increase (decrease) along with the surface roughness for a micro-pillar patterned (porous) surface, and if the surface roughness is fixed, a small droplet tends to ease the critical electrowetting condition for the transition. Besides, three dimensional phase diagrams in terms of electrowetting number, surface roughness, and droplet volume are constructed to illustrate the Cassie-Baxter-Wenzel wetting transition. Our theoretical model can be used to explain the previous experimental results about the Cassie-Baxter-Wenzel wetting transition reported in the literature.

\end{abstract}
\date{\today}

\maketitle

\section{INTRODUCTION}
The role of wettability is widely studied in order to gain highly water-repellent substrates referred to as superhydrophobic surfaces with an ultrahigh apparent contact angle, a much smaller contact angle hysteresis~\cite{D.Quere2005,D.Quere2008}, and hydrodynamic slip~\cite{C.Choi2006,P.Joseph2006,A.Steinberger2007}.
These properties rely on the nano- or the microscale topological structures of the surfaces which exhibit a broad range of applications in engineering such as self-cleaning~\cite{R.Blossey2003,K.M.Wisdom2013}, water proofing~\cite{A.Lafuma2003,M.Nosonovsky2007}, drag reduction~\cite{R.Truesdell2006}, anti-dew/reflection~\cite{J.B.Boreyko2009,R.H.Siddique2015}, bactericidal activity~\cite{E.P.Ivanova2012,E.P.Ivanova2013,X.L.Li2016,KeXiao2020}, and so on.
Typically, when a liquid droplet rests on such a roughened surface, a classical description of the droplet is characterized by Wenzel (W)~\cite{R.N.Wenzel1936} and Cassie-Baxter (CB)~\cite{A.B.D.Cassie1944} wetting states. The former one corresponds to a homogenous wetting state in which liquid penetrates into the texture, i.e. a fully wetted state losing superhydrophobicity, whereas the droplet in Cassie-Baxter state merely suspends on the tips of the rough surface, featured with air being trapped in the cavities of the rough surface topography.
Generally, the transition between these two states can be triggered via various approaches by tuning external control parameters, such as passive strategies that rely, for example, on the utilization of gravity force~\cite{Z.Yoshimitsu2002,B.Majhy2020}, evaporation~\cite{P.Tsai2010,X.M.Chen2012,H.C.M.Fernandes2015}, and Laplace pressure~\cite{A.Giacomello2012,P.Lv2014}, and active strategies that employ surface acoustic wave~\cite{A.Sudeepthi2020}, vibration~\cite{E.Bormashenko2007,J.B.Boreyko2009October}, and even electric field~\cite{G.Manukyan2011,J.M.Oh2011,R.Roy2018,B.X.Zhang2019}, a technique called electrowetting (EW).
The advantages of adaption to various geometries, little power consumption, and fast and precise fine-tuning of the wetting state make EW a prevalent technique receiving significant research interests in the past decade~\cite{F.Mugele2005,W.C.Nelson2012}.
It is well known that once an electric voltage is applied between a substrate and a droplet settling on it, the initial equilibrium contact angle of the droplet will be reduced to a new smaller value, leading to an alteration of its apparent wettability referred to as electrowetting-on-dielectric~\cite{F.Mugele2005,L.Q.Chen2014}.
 Such an EW phenomenon has attracted significant attention due to its extensive applications in lab-on-chip systems~\cite{F.Mugele2005} and microfluidic operations~\cite{M.G.Pollack2002,V.Bahadur2007}, and the discovery of the ability to induce wetting states transition~\cite{G.Manukyan2011,J.M.Oh2011,R.Roy2018,B.X.Zhang2019} and the droplet detachment~\cite{A.Cavalli2016,Q.Vo2019,Q.G.Wang2020,K.Xiao}.

Recently, numerous efforts using experiment~\cite{G.Manukyan2011,S.Berry2012,Y.Chen2019}, theoretical modeling~\cite{V.Bahadur2007,R.Roy2018}, and computer simulation~\cite{J.M.Oh2011,B.X.Zhang2019,A.M.Miqdad2016} have been devoted to getting a better understanding of the wetting transition triggered by electric field. By combining experiment and numerical simulation, it has been found that the stability of the CB state under EW is determined by the balance of the Maxwell stress and the Laplace stress~\cite{G.Manukyan2011}. Meanwhile, the wetting transition from CB state to W state is controlled by the energy barrier stemming from the pinning of the contact lines at the edges of the hydrophobic pillars~\cite{G.Manukyan2011}.
Based on a surface energy model, Roy et al. estimated the energy barriers for the EW-induced CB-to-W transition of a droplet on a mushroom-shaped re-entrant microstructures and an array of cylindrical microposts. They experimentally demonstrated that the transition on a mushroom structure is more resilient than that on an array of microposts~\cite{R.Roy2018}.
Besides, computer simulation also provides a useful complement to revealing the underlying mechanism of droplet wetting transition on textured surfaces.
By employing molecular dynamics simulations, Zhang et al.~\cite{B.X.Zhang2019} studied the mechanism behind the CB-to-W transition of a nanoscale water droplet resting on a nanogrooved surface under an external electric field, and found that there exists an energy barrier separating the CB state and the W state. In addition, they also discussed the dependence of the energy barrier on the electric filed strength, the groove aspect ratio, and the intrinsic contact angle of the groove.

Despite of the fact that the EW-induced transition have been extensively studied either via experimental or theoretical approaches, a systematic analytical understanding of the underlying mechanism, in particular, the dependence of critical electric voltage on surface roughness and droplet volume has not been explored.

In this paper, we establish a theoretical model to study the EW transition on micro-patterned surfaces through analyzing the difference of interfacial free energy between CB state and intermediate composite state.
The effects of surface roughness and droplet volume on the threshold voltage are discussed.
To further explore the interrelation among threshold voltage, surface roughness and droplet volume, three dimensional (3D) phase diagrams in the corresponding parameter space are also constructed.
We expect that our model can offer some guidance to the design and fabrication of the patterned surfaces and allow one to study EW transition on other types of patterned surfaces.

\section{THEORETICAL MODELING}
\begin{figure}[htp]
  \includegraphics[width=\linewidth,keepaspectratio]{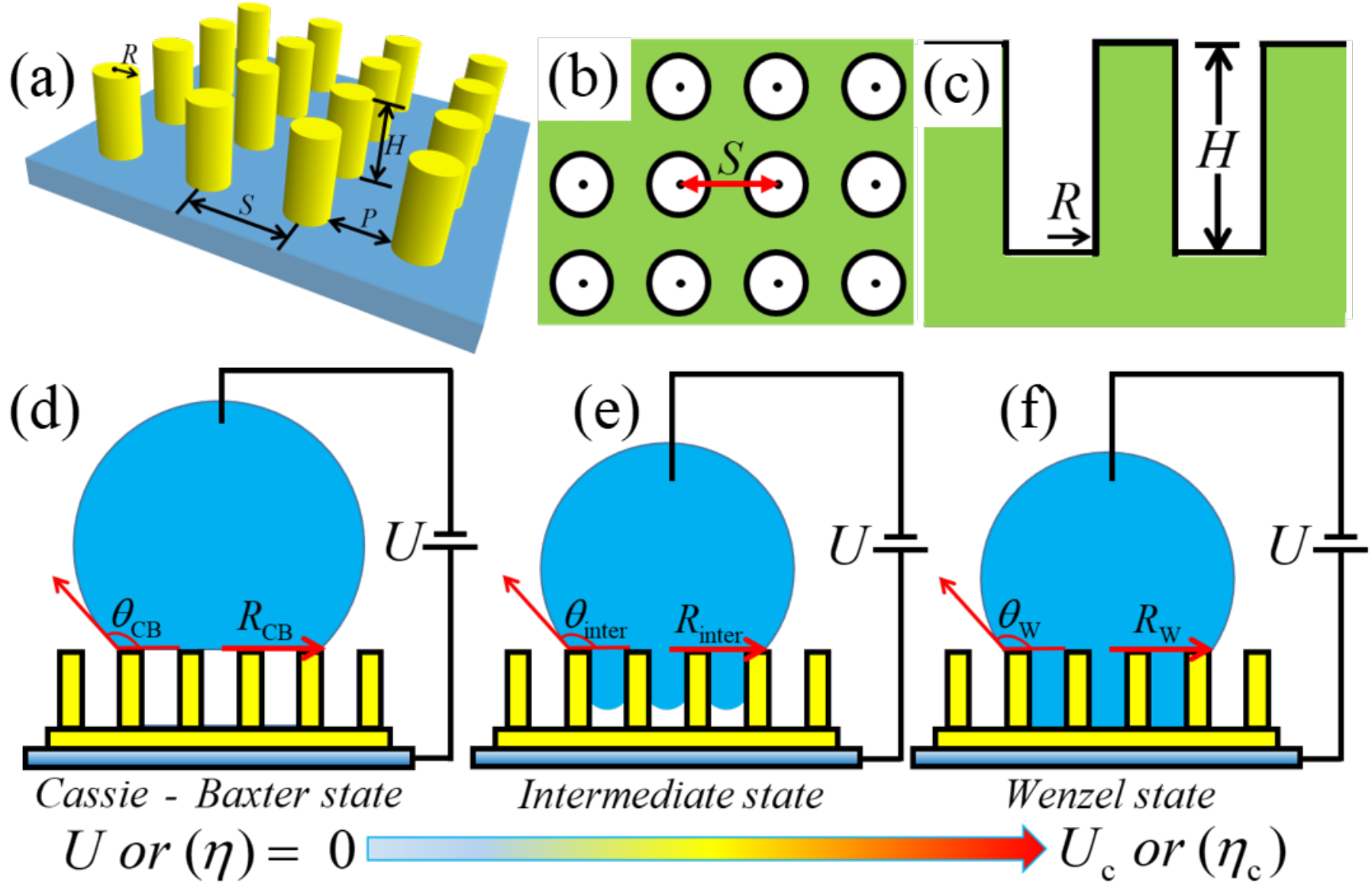}
  \caption{(Color online) (a) Schematic 3D picture of a microstructured surface with cylindrical pillars of radius $R$, height $H$ and gap pitch $P$ on a periodic square lattice with pillar-to-pillar spacing $S$. Schematic (b) top view and (c) side view of a regular array of pores of radius $R$, depth $H$, gap pitch $P$, and center-to-center pore spacing $S$. Different wetting states of a droplet on a microtextured surface in the presence of an external electric voltage: (d) CB state, (e) intermediate composite state, and (f) W state. \label{schematic}}
\end{figure}

We begin our investigation by considering an EW setup consisting of a millimeter-sized sessile water droplet deposited on two different types of superhydrophobic surface decorated with, respectively, a square lattice of cylindric mircropillars and a regular array of pores, as shown in Fig.~\ref{schematic}. The three dimensional (3D) geometry of the micro-pillar patterned surface in this paper is schematically shown in Fig.~\ref{schematic}(a). The top view and the side view of the squarely distributed micropore-patterned surface are sketched in Figs.~\ref{schematic}(b) and~\ref{schematic}(c) respectively. Here, the cylindrical pillars and the pores are characterized by their radius $R$, height $H$, and gap pitch $P$ (or center-to-center interspacing $S$) between neighboring pillars or pores, respectively. Traditionally, nondimensional parameters roughness factor $r$ and solid fraction $\phi_s$, which are defined as the ratio of the actual area of the solid surface to its projection area and the ratio of the contact solid surface (tip of the pillars or the pores) to the total horizontal surface respectively, are commonly used to represent the level of roughness for the textured surfaces. Geometrically, in this paper, the roughness factor is given by $r=1+2\pi RH/S^2$, and the solid fraction $\phi_s$ are written as $\pi R^2/S^2$ for the pillar patterned surface and $1-\pi R^2/S^2$ for the porous surface respectively. The two possible wetting states, i.e., CB state and W state, are illustrated by Figs.~\ref{schematic}(d) and ~\ref{schematic}(f), between which there exists an energy barrier, i.e., an intermediate composite state [see Fig.~\ref{schematic}(e)], which can be lowered to a level below CB state by applying an external electric voltage across the droplet.
This leads to a CB-W wetting transition.
In order to probe the critical electric voltage that triggers the wetting transition, it is necessary to calculate the total interfacial free energy for CB state and intermediate composite state, which, in general, is given by the sum of all the surface energies between the droplet and the patterned surface, i.e.,
\begin{align}
G=\gamma_{\rm lv}A_{\rm lv}+\gamma_{\rm sv}A_{\rm sv}+\gamma_{\rm ls}^{\rm eff}A_{\rm ls},\label{eq:TFE}
\end{align}
where $A_{\rm lv}$, $A_{\rm sv}$, and $A_{\rm ls}$ are the areas of the liquid-vapor, solid-vapor, and liquid-solid interfaces, and $\gamma_{\rm lv}$, $\gamma_{\rm sv}$, and $\gamma_{\rm ls}^{\rm eff}$ are the liquid-vapor, solid-vapor, and effective liquid-solid interfacial energy density, respectively. Here $\gamma_{\rm ls}^{\rm eff}=\gamma_{\rm ls}-\eta \gamma_{\rm lv}$ with $\eta=\varepsilon_{0}\varepsilon U^2/2d\gamma_{\rm lv}$ represents the dimensionless electrowetting number, where $\varepsilon_0$, $\varepsilon$, and $d$ are the dielectric permittivity in vacuum, the relative dielectric constant, and the thickness of the insulating layer, respectively.
Let $A_{\rm t}=A_{\rm sv}+A_{\rm ls}$, the total area of the solid surface including solid-vapor and liquid-solid interfaces, and with a consideration of the Young's equation $\gamma_{\rm sv}-\gamma_{\rm ls}=\gamma_{\rm lv}\cos{\theta_{\rm Y}}$, Eq.~(\ref{eq:TFE}) can be converted to
\begin{align}
G=\gamma_{\rm lv}A_{\rm lv}-\gamma_{\rm lv}(\cos{\theta_{\rm Y}}+\eta)A_{\rm ls}+\gamma_{\rm sv}A_{\rm t},\label{eq:TFESimplified}
\end{align}
where $\theta_{\rm Y}$ is the apparent contact angle at the equilibrium state.
Here we assume that the volume of the droplet is small, corresponding to a characteristic size smaller than the capillary length $l_{\rm c}=\sqrt{\gamma_{\rm lv}/\rho g}\sim 2.7~{\rm mm}$, where $\rho$ and $g$ are the density of the water and the gravitational acceleration respectively. In this case the gravitational effect can be neglected, and the shape of the droplet associated to the transition can be treated as a sphere. In particular, the total interfacial free energy of a water droplet in CB state on a patterned surface, as schematically illustrated by Fig.~\ref{schematic}(d), can be calculated as
\begin{align}
G_{\rm CB}=&\gamma_{\rm lv}\pi R_{\rm CB}^2\Big[\nu(\theta_{\rm CB})+(1-\phi_s)-(\cos{\theta_{\rm Y}}+\eta)\phi_s\Big] \notag\\
& +\gamma_{\rm sv}A_{\rm t},\label{eq:G_CB}
\end{align}
where $\nu(\theta)=2/(1+\cos\theta)$ is a dimensionless function. In addition, as the evaporation effect of the water droplet is excluded as well, it is reasonable to assume that the droplet volume $V_0$ is conserved. Therefore the base radius $R_{\rm CB}$ and the contact angle $\theta_{\rm CB}$ of the droplet can be determined according to a minimization of the global energy under the constraint of fixed droplet volume.
Similarly, the total interfacial free energy of the intermediate composite state reads
\begin{align}
G_{\rm inter}=&\gamma_{\rm lv}\pi R_{\rm inter}^2\Bigg\{\nu(\theta_{\rm inter})+(1-\phi_s)\nu(\theta_{\rm Y}-\frac{\pi}{2})- \notag\\
& (\cos{\theta_{\rm Y}}+\eta)\bigg[\phi_s+(r-1)\frac{h}{H}\bigg]\Bigg\}+\gamma_{\rm sv}A_{\rm t},\label{eq:G_mid}
\end{align}
where $R_{\rm inter}$ and $h$ are the base radius of the droplet and the penetration height from the tip of the pillars or pores to the point at which the curved interstitial liquid-vapor meniscus touches the edges of the walls (see Appendix for the calculation of $h$), respectively.
\begin{figure}[h]
 \includegraphics[width=\linewidth,keepaspectratio]{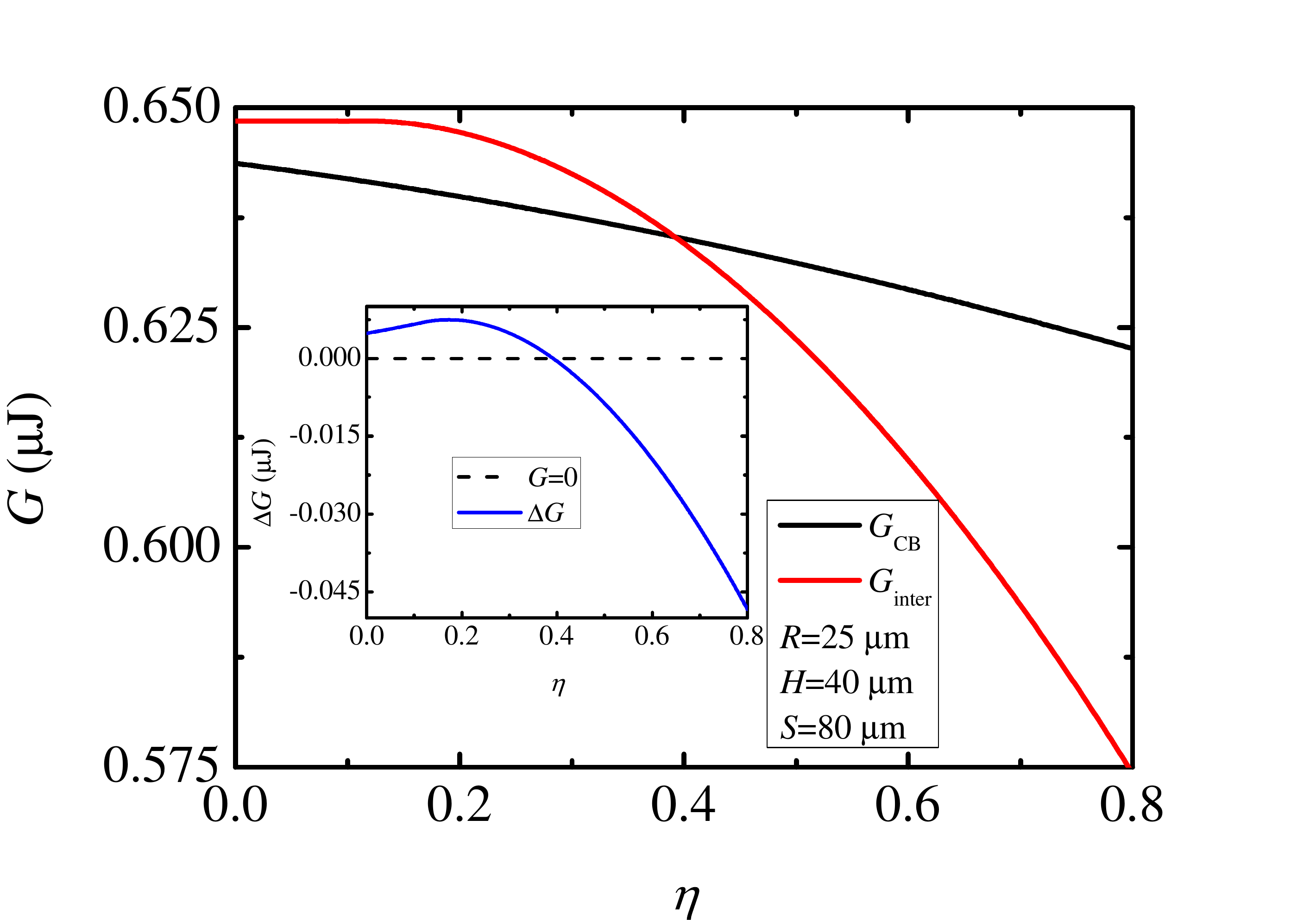}
  \caption{(Color online) Representative total interfacial free energies of the CB and intermediate states of a droplet on a micropillar-patterned surface: $G_{\rm CB}$ and $G_{\rm inter}$, respectively, as a function of EW number. The inset shows their difference, $\Delta G=G_{\rm inter}-G_{\rm CB}$, which gives the critical EW number (critical external electric voltage) for the wetting transition when $\Delta G=0$. \label{GVseta}}
\end{figure}

\section{RESULTS AND DISCUSSION}
In our model, we take into account all the interfacial free energies for a CB state and an intermediate composite state of a three-dimensional droplet when placed on a micropillar- or a pore- patterned surface. Our calculations in this paper were carried out by using $\gamma_{\rm lv}=72.8~{\rm mN\cdot m^{-1}}$, $\varepsilon=3.2$, $d=3~\mu{\rm m}$ and $\theta_{\rm Y}=115^{\circ}$, respectively~\cite{R.Roy2018}. Without the application of an external field, as valued by the intercept in Fig.~\ref{GVseta}, the energy of the intermediate composite state is higher than that of the CB state, denying the occurrence of a CB-W transition. However, once an external field is applied and increased, it is found that although the energy of the CB state decreases, the energy of the intermediate composite state decreases in a steady and more rapid way, and there exists an intersection point where the droplet in CB state crosses the energy barrier created by the intermediate composite state and changes to a Wenzel state, as shown in Fig.~\ref{GVseta}. Such a critical condition corresponds to a critical voltage (or a critical EW number $\eta_{\rm c}$) which can be estimated by equating the energies of these two states for a given textured surface.
\begin{figure}[htp]
  \includegraphics[width=\linewidth,keepaspectratio]{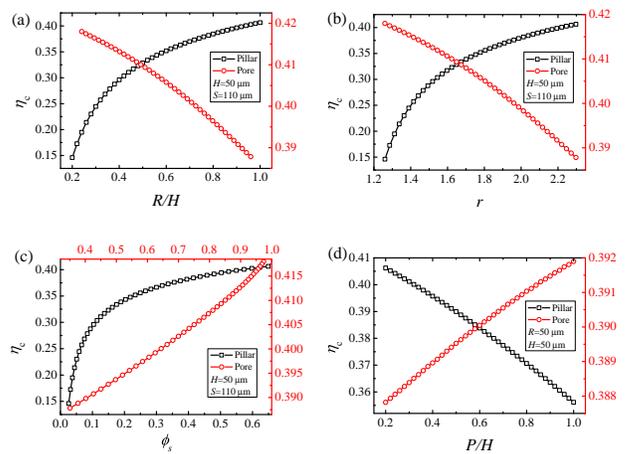}
  \caption{(Color online) The critical EW number $\eta_c$ for the CB-W transition vs (a) aspect ratio $R/H$, (b) surface roughness $r$, (c) solid fraction $\phi_s$, and (d) $P/H$, where the black square dot curve and the red circle dot curve stand for micropillar- and pore- patterned surfaces, respectively. \label{RSVsetac}}
\end{figure}
For a fixed droplet volume, the EW number $\eta_{\rm c}$ is found to remarkably depend on the geometric features of the structured surface (surface roughness), such as aspect ratio $R/H$, relative pitch $P/H$ (density $1/S^2=1/(P+2R)^2$), surface roughness $r$ and solid fraction $\phi_{\rm s}$, as shown in Fig.~\ref{RSVsetac}.

 It can be seen that, as shown by the black square dot curve in Fig.~\ref{RSVsetac}(a), it becomes harder (easier) for a CB-W transition on a micropillar-patterned surface (porous surface) to occur with the increase of its aspect ratio $R/H$. An alternative illustration of Fig.~\ref{RSVsetac}(a) is to replace the dimensionless aspect ratio $R/H$ by surface roughness $r$, indicating that roughness suppresses (enhances) EW-induced CB-W transition on a micropillar-patterned surface (porous surface) (Fig.~\ref{RSVsetac}(b)). Besides, as the aspect ratio also correlates with solid fraction $\phi_{\rm s}$, it is possible to depict the critical EW number in terms of solid fraction $\phi_{\rm s}$ for both pillar- and pore- patterned surfaces, as shown in Fig.~\ref{RSVsetac}(c).
Apart from the aspect ratio, the distribution density of pillars and pores also plays an important role in determining the critical EW number. A more deeper investigation, as shown in Fig.~\ref{RSVsetac}(d), exhibits that a reduction of critical EW number $\eta_{\rm c}$ can be achieved by increasing (decreasing) the pitch of pillar-patterned (pore-patterned) surfaces.
\begin{figure}[htp]
  \includegraphics[width=\linewidth,keepaspectratio]{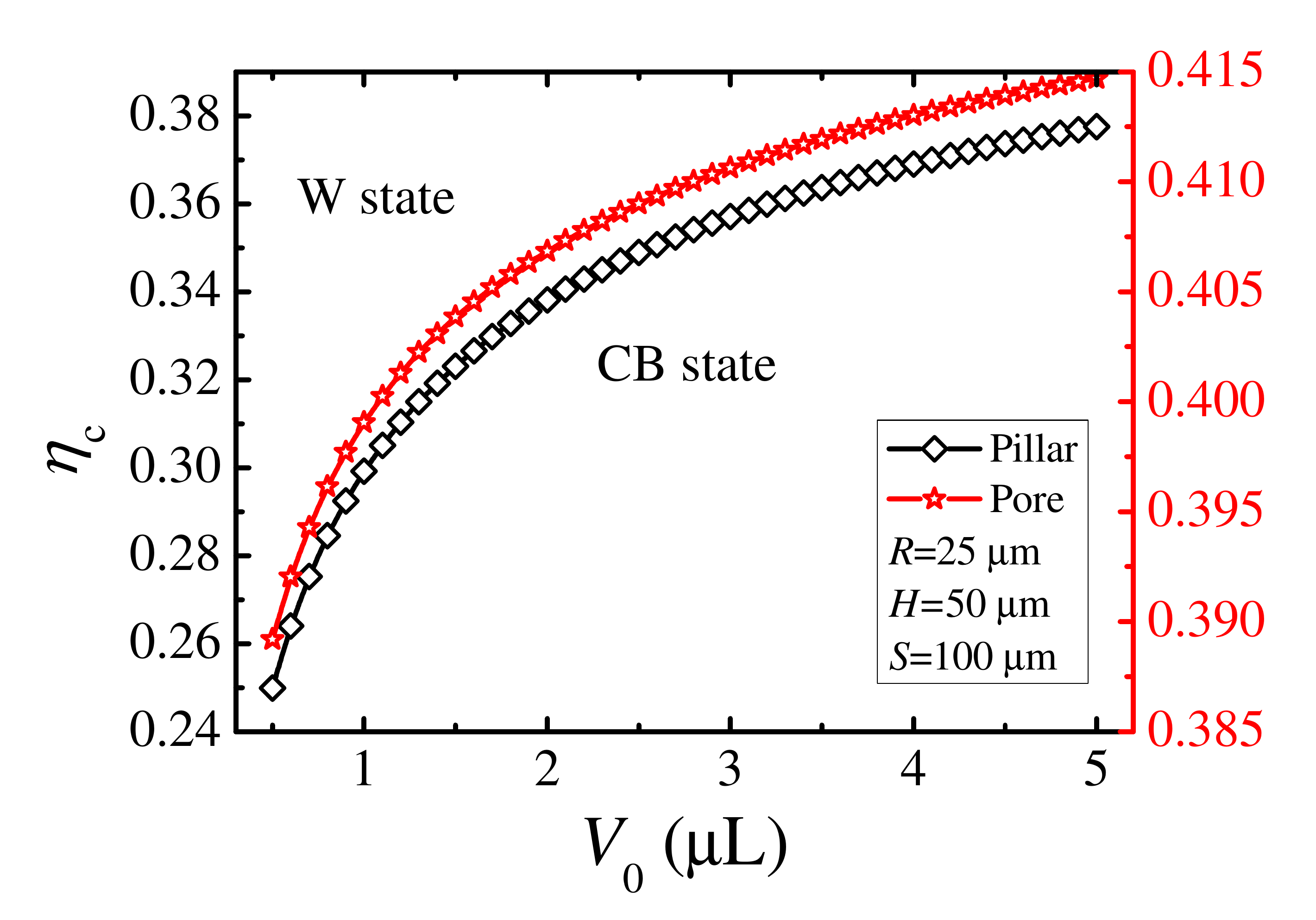}
  \caption{(Color online) The critical EW number $\eta_c$ for the CB-W transition vs droplet volume $V_0$, where the black square dot curve and the red circle dot curve represent micropillar- and pore- patterned surfaces, respectively. \label{VoVsetac}}
\end{figure}

It has been found experimentally that the onset of CB-W wetting transition occurs at a certain droplet size as the droplet gets smaller during evaporation process~\cite{P.Tsai2010,X.M.Chen2012,H.C.M.Fernandes2015}. Meanwhile, the thermodynamic favorable wetting
state also depends on droplet volume~\cite{K.Xiao2017}. Such a phenomenon can be explained by our theoretical model. Figure~\ref{VoVsetac} exhibits the effect of droplet volume $V_0$ on wetting for a set of fixed surface geometric parameters ($R=25~{\rm \mu m}$, $H=50~{\rm \mu m}$, and $S=100~{\rm \mu m}$), revealing that the larger the droplet size, the higher the critical voltage for the wetting transition to occur.

\begin{figure}[htp]
  \includegraphics[width=\linewidth,keepaspectratio]{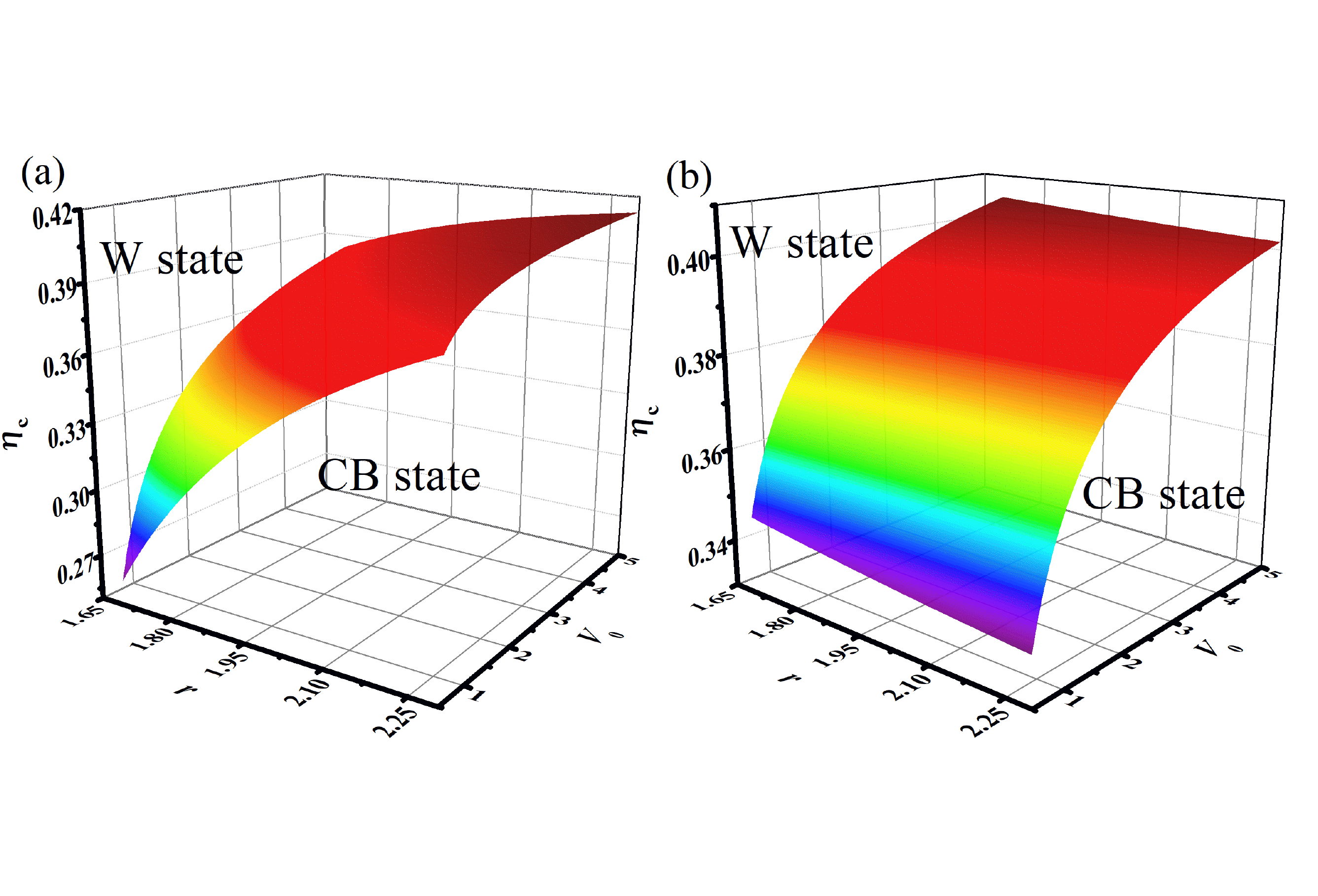}
  \caption{(Color online) 3D phase diagrams in terms of the critical EW number $\eta_c$, surface roughness $r$ and droplet volume $V_0$ for (a) a micropillar-patterned surface and (b) a pore-patterned surface. \label{phasediagramRoughnessVsVo}}
\end{figure}

Finally, in order to comprehensively understand how surface roughness and droplet size affect the critical value of $\eta_c$ as a whole, 3D phase diagrams for micropillar and pore-patterned surface are constructed in terms of EW number, surfce roughness, and droplet volume, as demonstrated in Fig.~\ref{phasediagramRoughnessVsVo}. It is found that all phase diagrams are divided into two regimes, namely CB state (under the curved surface) and W state (above the curved surface) separated by a coexisting curved surface representing the critical condition for the wetting transition.
According to the 3D phase diagrams, we can deduce that a higher (lower) critical EW number $\eta_c$ is required to trigger the CB-W transition for rougher pillar- (pore-) patterned surfaces. While for large droplet, higher $\eta_c$ is needed for the wetting transition regardless of the geometric pattern of the surfaces. Therefore, it becomes possible that the CB-W transition triggered by EW effect can be effectively inhibited by engineering a surface with hierarchical roughness or by adopting large droplet.

To examine the validity of our theoretical model, it is necessary to compare our theoretical predictions with experimental results.
For example, Bahadur et al.~\cite{V.Bahadur2008} showed via experiments that the observed transition voltage is 35~V for microstructured surface with roughness 2.87, solid fraction 0.23, and pillar height 43.1~${\rm \mu m}$, while the transition voltage has to be increased to 58~V to observe the wetting transition for the microstructured surface with same pillar height but an increased surface roughness 3.71 and solid fraction 0.55, a result in good agreement with our present conclusions.
In addition, it has been found by the experimental observations made by Manukyan et al.~\cite{G.Manukyan2011} that the critical voltage causing the cavities filled with water and the lateral propagation decreases as the gap widths between the pillars were widened (equivalent to a reduction of surface roughness and solid fraction), a conclusion also in accordance with our results.
What's more, Roy et al.~\cite{R.Roy2018} also reported that the more sparsely the cylindrical microposts are distributed, the higher the EW voltage is required to trigger the CB-W transition. These results all support the theoretical model proposed in this paper.

\section{CONCLUSION}
In this paper we developed a model to interpret the CB-W wetting transition triggered by an external electric field for a three-dimensional water droplet deposited on a micropillar- or a pore-patterned surface. It is found that the electric field lowers the energy barrier created by the intermediate composite state, which allows the droplet initially in CS state to cross and complete the EW-induced CB-W transition.
 The critical value of the electric field applied for the transition is influenced by the geometrical parameters for the thermodynamic wetting states, such as the base radius and apparent contact angle of a droplet, and the critical voltage for the CB-W wetting transition. 3D phase diagrams in terms of EW number, surface roughness, and droplet volume are constructed. It is shown that low (high) roughness, low (high) pitch, small (small) solid fraction, and small (small) droplet size encourages a CB-W wetting transition triggered by an external field, a conclusion in good agreement with previous investigations reported in the literature.

${\dag}$ These authors contributed equally to this work.

\begin{acknowledgements}
This work was funded by the National Science Foundation of China under Grant No. 11974292 and No. 11947401.
\end{acknowledgements}

\section{Appendix: The calculation of the base radius and the penetration height of a droplet placed on a patterned surface}
For all the wetting states in the main text, the base radius of a droplet is determined under the constraint of fixed volume, which, in CB state, can be written as
\begin{align}
V_{\rm CB}=\frac{\pi}{3}R_{\rm CB}^3\mu(\theta_{\rm CB})=V_0,\label{eq:V_CB_pillar}
\end{align}
where $\mu(\theta)=(2+\cos\theta)(1-\cos\theta)^2/\sin^3\theta$ is a dimensionless function, and $R_{\rm CB}$ and $\theta_{\rm CB}$ are the base radius and the apparent contact angle of the droplet respectively. Such an equation can also be rewritten as
\begin{align}
R_{\rm CB}=\bigg[\frac{3V_0}{\pi\mu(\theta_{\rm CB})}\bigg]^{1/3}.\label{eq:R_CB_pillar}
\end{align}
When a droplet on a microstructured surface reaches the intermediate state, its volume can be divided into two parts, i.e., the one on the top of the patterned surface and the other penetrating into the interspacing of the pillars.
The volume of the spherical cap above the patterned surface is given by
\begin{align}
V_{\rm top}^{\rm inter}=\frac{\pi}{3}R_{\rm inter}^3\mu(\theta_{\rm inter}).\label{eq:V_top_mid_pillar}
\end{align}
Strictly speaking, the equilibrium configuration of the curved liquid-vapor interface in a unit cell between the pillars needs to be determined by the Young-Laplace equation. However, due to the fact that the droplet volume filling the space around the pillars is much smaller than that of the spherical cap above the patterned surface, it is reasonable to treat the shape of the liquid-vapor interface in a unit cell between the pillars as a spherical cap with effective capillary radius $R_{\rm cap}^{\rm eff}$ defined by~\cite{R.Roy2018}
\begin{align}
R_{\rm cap}^{\rm eff}=S\bigg(\frac{1-\phi_s}{\pi}\bigg)^{1/2}.\label{eq:R_cap_pillar}
\end{align}
Thus, the height of the corresponding spherical cap is calculated as
\begin{align}
h_{\rm cap}^{\rm eff}=R_{\rm cap}^{\rm eff}\frac{1-\cos\big(\theta_{\rm Y}-\frac{\pi}{2}\big)}{\sin\big(\theta_{\rm Y}-\frac{\pi}{2}\big)}.\label{eq:h_cap_pillar}
\end{align}
The penetration height $h$ in the main text can be obtained as $h=H-h_{\rm cap}^{\rm eff}$.
Then the droplet volume underneath the top spherical cap corresponding to the volume penetrated into the interspacing of the pillars is given by
\begin{align}
V_{\rm bottom}^{\rm inter}=\frac{\pi}{3}R_{\rm inter}^2(1-\phi_s)\bigg[3h+R_{\rm cap}^{\rm eff}\mu\big(\theta_{\rm Y}-\frac{\pi}{2}\big)\bigg].\label{eq:V_bottom_mid_pillar}
\end{align}
Given this, the base radius $R_{\rm mid}$ of a droplet in the intermediate state can be found by solving the following equation
\begin{align}
V_0=\frac{\pi}{3}R_{\rm inter}^3\mu(\theta_{\rm inter})+\frac{\pi}{3}R_{\rm inter}^2(1-\phi_s)\bigg[3h+R_{\rm cap}^{\rm eff}\mu\big(\theta_{\rm Y}-\frac{\pi}{2}\big)\bigg].\label{eq:V}
\end{align}
Similarly, when a droplet on a pore-patterned surface gets into intermediate state, the calculation of its base radius can be done in the same manner as above so long as we replace the effective capillary radius of the spherical cap of the bottom part by the radius of the pore $R$.


\begin{thebibliography}{99}
\bibitem{D.Quere2005} D. Qu\'{e}r\'{e}, Non-sticking drops, Rep. Prog. Phys. \textbf{68}, 2495 (2005).
\bibitem{D.Quere2008} D. Qu\'{e}r\'{e}, Wetting and Roughness, Annu. Rev. Mater. Res. \textbf{38}, 71 (2008).
\bibitem{C.Choi2006} C. Choi and C. J. Kim, Large Slip of Aqueous Liquid Flow over a Nanoengineered Superhydrophobic Surface, Phys. Rev. Lett. \textbf{96}, 066001 (2006).
\bibitem{P.Joseph2006} P. Joseph, C. Cottin-Bizonne, J.-M. Beno\^{\i}t, C. Ybert, C. Journet, P. Tabeling, and L. Bocquet, Slippage of Water Past Superhydrophobic Carbon Nanotube Forests in Microchannels, Phys. Rev. Lett. \textbf{97}, 156104 (2006).
\bibitem{A.Steinberger2007} A. Steinberger, C. Cottin-Bizonne, P. Kleimann, and E. Charlaix, High friction on a bubble mattress, Nature Mater. \textbf{6}, 665 (2007).
\bibitem{R.Blossey2003} R. Blossey, Self-Cleaning Surfaces Virtual Realities, Nat. Mater. \textbf{2} 301-306 (2003)
\bibitem{K.M.Wisdom2013} K. M. Wisdom, J. A. Watson, X. Qu, F. Liu, G. S. Watson, and C. H. Chen, Selfcleaning of superhydrophobic surfaces by self-propelled jumping condensate, Proc. Natl. Acad. Sci. U. S. A. {\bf 20}, 7992-7997 (2013).
\bibitem{A.Lafuma2003} A. Lafuma and D. Quere, Superhydrophobic States, Nat. Mater., \textbf{2} 457-460 (2003).
\bibitem{M.Nosonovsky2007} M. Nosonovsky and B. Bhushan, Biomimetic Superhydrophobic Surfaces: Multiscale Approach, Nano Lett. \textbf{7}, 2633 (2007).
\bibitem{R.Truesdell2006} R. Truesdell, A. Mammoli, P. Vorobieff, F. van Swol, and C. J. Brinker, Drag reduction on a patterned superhydrophobic surface, Phys. Rev. Lett. \textbf{97}, 044504 (2006).
\bibitem{J.B.Boreyko2009} J. B. Boreyko and C. H. Chen, Self-propelled dropwise condensate on superhydrophobic surfaces, Phys. Rev. Lett. {\bf 103}, 184501 (2009).
\bibitem{R.H.Siddique2015} R. H. Siddique, G. Gomard, and H. Holscher, The role of random nanostructures for the omnidirectional anti-reflection properties of the glasswing butterfly, Nat. Commun. \textbf{6}, 301 (2015).
\bibitem{E.P.Ivanova2012} E. P. Ivanova, J. Hasan, H. K. Webb, V. K. Truong, G. S. Watson, J. A. Watson, V. A. Baulin, S. Pogodin, J. Y. Wang, M. J. Tobin, C. Lobbe, and R. J. Crawford, Natural Bactericidal Surfaces: Mechanical Rupture of Pseudomonas aeruginosa Cells by Cicada Wings, Small \textbf{8}, 2489 (2012).
\bibitem{E.P.Ivanova2013} E. P. Ivanova, J. Hasan, H. K. Webb, G. Gervinskas, S. Juodkazis, V. K. Truong, A. H. F. Wu, R. N. Lamb, V. A. Baulin, G. S. Watson, J. A. Watson, D. E. Mainwaring, and R. J. Crawford, Bactericidal activity of black silicon, Nat. Commun. \textbf{4}, 359 (2013).
\bibitem{X.L.Li2016} X. L. Li, Bactericidal mechanism of nanopatterned surfaces, Phys. Chem. Chem. Phys. \textbf{18}, 1311-1316 (2016).
\bibitem{KeXiao2020} K. Xiao, X. Z. Cao, X. Chen, H. Z. Hu, and C. X. Wu, Bactericidal efficacy of nanopatterned surface tuned by topography, J. Appl. Phys. \textbf{128}, 064701 (2020).
\bibitem{R.N.Wenzel1936} R. N. Wenzel, Resistance of Solid Surfaces to Wetting by Water, Ind. Eng. Chem. \textbf{28}, 988-994 (1936).
\bibitem{A.B.D.Cassie1944} A. B. D. Cassie and S. Baxter, Wettability of Porous Surface, Trans. Faraday Soc. \textbf{40}, 546-551 (1944).
\bibitem{Z.Yoshimitsu2002} Z. Yoshimitsu, A. Nakajima, T. Watanabe, and K. Hashimoto, Effects of Surface Structure on the Hydrophobicity and Sliding Behavior of Water Droplets, Langmuir \textbf{18}, 5818 (2002).
\bibitem{B.Majhy2020} B. Majhy, V.P. Singh, A. K. Sen, Understanding wetting dynamics and stability of aqueous droplet over superhydrophilic spot surrounded by superhydrophobic surface, J. Colloid Interface Sci. \textbf{565}, 582-591 (2020).
\bibitem{P.Tsai2010} P. Tsai, R. G. H. Lammertink, M. Wessling, and D. Lohse, Evaporation-Triggered Wetting Transition for Water Droplets upon Hydrophobic Microstructures, Phys. Rev. Lett. \textbf{104}, 116102 (2010).
\bibitem{X.M.Chen2012} X. M. Chen, R. Y. Ma, J. T. Li, C. L. Hao, W. Guo, B. L. Luk, S. C. Li, S. H. Yao, and Z. K. Wang, Evaporation of Droplets on Superhydrophobic Surfaces: Surface Roughness and Small Droplet Size Effects, Phys. Rev. Lett. \textbf{109}, 116101 (2012).
\bibitem{H.C.M.Fernandes2015} H. C. M. Fernandes, M. H. Vainstein, and C. Brito, Modeling of Droplet Evaporation on Superhydrophobic Surfaces, Langmuir \textbf{31}, 7652-7659 (2015).
\bibitem{A.Giacomello2012} A. Giacomello, M. Chinappi, S. Meloni, and C. M. Casciola, Metastable Wetting on Superhydrophobic Surfaces: Continuum and Atomistic Views of the Cassie-Baxter-Wenzel Transition, Phys. Rev. Lett. \textbf{109}, 226102 (2012).
\bibitem{P.Lv2014} P. Lv, Y. Xue, Y. Shi, H. Lin, and H. Duan, Metastable States and Wetting Transition of Submerged Superhydrophobic Structures, Phys. Rev. Lett. \textbf{112}, 196101 (2014).
\bibitem{A.Sudeepthi2020} A. Sudeepthi, L. Yeo, and A. K. Sen, Cassie-Wenzel wetting transition on nanostructured superhydrophobic surfaces induced by surface acoustic waves, Appl. Phys. Lett. \textbf{116}, 093704 (2020).
\bibitem{E.Bormashenko2007} E. Bormashenko, R. Pogreb, G. Whyman, Y. Bormashenko, and M. Erlich, Vibration-induced Cassie-Wenzel wetting transition on rough surfaces, Appl. Phys. Lett. \textbf{90}, 201917 (2007).
\bibitem{J.B.Boreyko2009October} J. B. Boreyko and C. H. Chen, Restoring Superhydrophobicity of Lotus Leaves with Vibration-Induced Dewetting, Phys. Rev. Lett. \textbf{103}, 174502 (2009).
\bibitem{G.Manukyan2011} G. Manukyan, J. M. Oh, D. van den Ende, R. G. H. Lammertink, and F. Mugele, Electrical Switching of Wetting States on Superhydrophobic Surfaces: A Route Towards Reversible Cassie-to-Wenzel Transitions, Phys. Rev. Lett. 106, 014501 (2011).
\bibitem{J.M.Oh2011} J. M. Oh, G. Manukyan, D. van den Ende, and F. Mugele, Electric-field-driven instabilities on superhydrophobic surfaces, EPL \textbf{93}, 56001 (2011).
\bibitem{R.Roy2018} R. Roy, J. A. Weibel, and S. V. Garimella, Re-entrant Cavities Enhance Resilience to the Cassie-to-Wenzel State Transition on Superhydrophobic Surfaces during Electrowetting, Langmuir \textbf{34}, 12787-12793 (2018).
\bibitem{B.X.Zhang2019} B. X. Zhang, S. L. Wang, and X. D. Wang, Wetting Transition from the Cassie-Baxter State to the Wenzel State on Regularly Nanostructured Surfaces Induced by an Electric Field, Langmuir \textbf{35}, 662-670 (2019).
\bibitem{F.Mugele2005} F. Mugele and J. C. Baret, Electrowetting: from basics to applications, J. Phys.: Condens Matter {\bf 17}, R705-R774 (2005).
\bibitem{W.C.Nelson2012} W. C. Nelson and C. J. Kim, Droplet Actuation by Electrowetting-on-Dielectric (EWOD): A Review, J. Adhes. Sci. Technol. {\bf 26}, 1747-1771 (2012).
\bibitem{L.Q.Chen2014} L. Q. Chen,E. Bonaccurso, Electrowetting - From statics to dynamics, Adv. Colloid Interface Sci. {\bf 210}, 2-12 (2014).
\bibitem{M.G.Pollack2002} M. G. Pollack, A. D. Shenderov, and R. B. Fair, Electrowetting-based actuation of droplets for integrated microfluidics, Lab Chip \textbf{2}, 96-101 (2002).
\bibitem{V.Bahadur2007} V. Bahadur and S. V. Garimella, Electrowetting-Based Control of Static Droplet States on Rough Surfaces, Langmuir \textbf{23}, 4918-4924 (2007).
\bibitem{A.Cavalli2016} A. Cavalli, D. J. Preston, E. Tio, D. W. Martin, N. Miljkovic, E. N. Wang, F. Blanchette, and J. W. M. Bush, Electrically induced drop
detachment and ejection, Phys. Fluids {\bf 28}, 022101 (2016).
\bibitem{Q.Vo2019} Q. Vo and T. Tran, Critical Conditions for Jumping Droplets, Phys. Rev. Lett. {\bf 123}, 024502 (2019).
\bibitem{Q.G.Wang2020} Q. G. Wang, M. Xu, C. Wang, J. P. Gu, N. Hu, J. F. Lyu, and W. Yao, Actuation of a Nonconductive Droplet in an Aqueous Fluid by
Reversed Electrowetting Effect, Langmuir {\bf 36}, 8152-8164 (2020).
\bibitem{K.Xiao} K. Xiao and C. X. Wu, Critical condition for electrowetting-induced detachment of a droplet from a curved surface, arXiv:2012.07255 (2020).
\bibitem{S.Berry2012} S. Berry, T. Fedynyshyn, L. Parameswaran, and A. Cabral, Switchable electrowetting of droplets on dual-scale structured surfaces, J. Vac. Sci. Technol. B \textbf{30}, 06F801 (2012).
\bibitem{Y.Chen2019} Y. Chen, Y. Suzuki, and K. Morimoto, Electrowetting-Dominated Instability of Cassie Droplets on Superlyophobic Pillared Surfaces, Langmuir \textbf{35}, 2013-2022 (2019).
\bibitem{A.M.Miqdad2016} A. M. Miqdad, S. Datta, A. K. Das, and P. K. Das, Effect of electrostatic incitation on the wetting mode of a nano-drop over a pillar-arrayed surface, RSC Adv. \textbf{6}, 110127 (2016).
\bibitem{K.Xiao2017} K. Xiao, Y. P. Zhao, G. Ouyang, and X. L. Li, An analytical model of nanopatterned superhydrophobic surfaces, J. Coat. Technol. Res. \textbf{14}, 1297-1306 (2017).
\bibitem{V.Bahadur2008} V. Bahadur and S. V. Garimella, Electrowetting-Based Control of Droplet Transition and Morphology on Artificially Microstructured Surfaces, Langmuir \textbf{24}, 8338-8345 (2008).
\end{thebibliography}

\end{document}